\documentclass[11pt]{IEEEtran}

\ifCLASSOPTIONcompsoc
  \usepackage[caption=false,font=normalsize,labelfont=sf,textfont=sf]{subfig}
\else
  \usepackage[caption=false,font=footnotesize]{subfig}
\fi

\IEEEoverridecommandlockouts                              

\usepackage{amsmath}
\usepackage{amssymb}
\usepackage{algorithmic}
\usepackage{algorithm}
\usepackage{color}
\usepackage{graphicx}
\usepackage{cite}
\usepackage{array}
\usepackage{subfloat}
\usepackage{multirow}
\usepackage{booktabs}
\usepackage{xcolor}
\usepackage{hyperref}

\newcommand{\sign}{\mathop{\bf sign}}

\newcommand{\vct}[1]{\boldsymbol{#1}}
\newcommand{\mtx}[1]{\boldsymbol{#1}}
\newcommand{\va}{\vct{a}}

\newcommand{\vd}{\vct{d}}

\newcommand{\vm}{\vct{m}}

\newcommand{\vq}{\vct{q}}

\newcommand{\vu}{\vct{u}}

\newcommand{\vx}{\vct{x}}

\newcommand{\valpha}{\vct{\alpha}}

\newcommand{\vxi}{\vct{\xi}}

\newcommand{\vone}{\vct{1}}

\newcommand{\mB}{\mtx{B}}

\newcommand{\mD}{\mtx{D}}

\newcommand{\mH}{\mtx{H}}

\newcommand{\mK}{\mtx{K}}
\newcommand{\mL}{\mtx{L}}

\renewcommand{\sl}{s_\ell}

\hypersetup{
  colorlinks,
  citecolor=violet,
  linkcolor=red,
  urlcolor=blue}

\begin{document}
\title{Salt Reconstruction in Full Waveform Inversion with a Parametric Level-Set Method}

\author{Ajinkya~Kadu, Tristan~van~Leeuwen and Wim~A.~Mulder
\thanks{Ajinkya Kadu and Tristan van Leeuwen are with the Mathematical Institute, Utrecht University, The Netherlands. Wim Mulder is associated with Delft University of Technology, The Netherlands and Shell Global Solutions International B.V.
 }% <-this % stops a space
 }
% \thanks{Manuscript received April 19, 2005; revised August 26, 2015.}}

\maketitle

\begin{abstract}
Seismic full-waveform inversion tries to estimate subsurface medium parameters from seismic data. Areas with subsurface salt bodies are of particular interest because they often have hydrocarbon reservoirs on their sides or underneath. Accurate reconstruction of their geometry is a challenge for current techniques. This paper presents a parametric level-set method for the reconstruction of salt-bodies in seismic full-waveform inversion. We split the subsurface model in two parts: a background velocity model and the salt body with known velocity but undetermined shape. The salt geometry is represented by a level-set function that evolves during the inversion. We choose radial basis functions to represent the level-set function, leading to an optimization problem with a modest number of parameters. A common problem with level-set methods is to fine tune the width of the level-set boundary for optimal sensitivity. We propose a robust algorithm that dynamically adapts the width of the level-set boundary to ensure faster convergence. Tests on a suite of idealized salt geometries show that the proposed method is stable against a modest amount of noise. We also extend the method to joint inversion of both the background velocity model and the salt-geometry.

The code to perform parametric level-set method for inversion is available at \url{https://github.com/ajinkyakadu/ParametricLevelSet}.
\end{abstract}

\section{Introduction}
Seismic imaging attempts to obtain detailed images of the subsurface from seismic data. Such data are obtained by placing explosive sources on or near the surface and recording the response with a large array of receivers at the surface. By repeating the experiment many times for varying source position, enough data can be gathered to form useful images. The scale of these experiments varies from tens of meters, for instance, for near-surface imaging, to tens of kilometers in oil and gas exploration, up to the whole Earth in global seismology. For hydrocarbon exploration, depths typically extend to several kilometers.

One of the main challenges of seismic imaging is that the propagation velocity of the seismic waves traveling through the subsurface is unknown. Since this velocity can vary significantly, both laterally and with depth, it has to be estimated prior to applying a conventional imaging method. % based on correlating back-propagated seismic data to forward-propagated source signals. 
A wrong subsurface velocity can lead to severely distorted images. The conventional workflow for seismic imaging, therefore, is to first estimate the subsurface velocity and subsequently back-propagate the data to obtain an image. The success of this two-step approach relies crucially on the ability to separate propagation and reflection effects in the data. This can only be done if the earth structure is relatively simple, e.g. smoothly varying with small perturbations. 
A particularly relevant setting in which the separation-of-scales argument fails to hold is in the presence of strong contrasts, such as salt diapirs, salt slabs, anhydrite or basalt layers. Salt geometries are of particular interest because they often have hydrocarbon reservoirs on their sides or underneath.

Full-waveform inversion (FWI) attempts to fit the data % \rmrk{at once? not if you go from low to high frequencies} 
using a fairly precise numerical model for wave-propagation in heterogeneous media. By posing the inverse problem as a non-linear least-squares problem, the velocity structure in the subsurface can, in principle, be estimated quantitatively. However, this optimization problem is severely non-linear and ill-posed, making it very difficult to obtain reasonable results without a good initial guess of the velocity parameters. Starting from an initial guess that is far away from the truth often leads to an incorrect subsurface model. While many approaches have been proposed to mitigate this problem, the issue remains unsolved.

In this paper, we aim to alleviate the ill-posedness of the problem to some extent by adding an appropriate regularization. In the particular setting of salt bodies, it is reasonable to assume that the subsurface can be described as one or more continuous bodies (salt) with known constant material parameters, surrounded by continuously varying parameters (sediment). Our approach is based on a level-set method, where we explicitly represent the shape of the salt body with a level-set function. By expanding the level-set function in some basis, we greatly reduce the effective number of parameters and obtain a non-linear optimization problem that is better behaved than the original non-linear least-squares problem. 

The paper is organized as follows. In Section \ref{fwi}, we review the details of the classical FWI approach and give an overview regularization approaches specifically aimed at the reconstruction of salt bodies. In Section \ref{levelset}, we present the basics of the parametric level-set method and discuss several practical issues in detail. 
Numerical results on a stylized seismic example are presented in Section \ref{results}. Finally, Section \ref{conclusions} concludes the paper.

\section{Full Waveform Inversion}
\label{fwi}
Full waveform inversion (FWI) is non-linear data-fitting scheme aiming to retrieve detailed estimates of subsurface properties from seismic data. 
The basic workflow of FWI is as follows: \emph{(1)} predict the observed data by solving a wave-equation, given an initial guess of the subsurface velocity, \emph{(2)} compute the difference between predicted and observed data, and \emph{(3)} update the velocity in order to improve the data fit.
This process is repeated in an iterative fashion until the residual drops below some tolerance.
An excellent overview of various flavors of this basic scheme can be found in \cite{virieux2009overview}.

There are several ways to model seismic wave propagation, We refer the reader to the reviews by \cite{carcione_seismic_2002} and \cite{kanao_modelling_2012}. For our purpose, it suffices to consider a two-dimensional scalar Helmholtz equation that models the acoustic pressure:

\begin{equation}
\label{eq:waveequation}
\omega^2 m(\vx)u(\omega,\vx) + \nabla^2 u(\omega,\vx) = q(\omega,\vx),
\end{equation}

where  $\vx\in \mathbb{R}^2$ denotes the subsurface position, $m(\vx)$ is the squared slowness with units $\text{s}^2/\text{m}^2$, $\omega$ is the angular frequency, $q$ the source term and $u$ the pressure wavefield. We consider an unbounded domain and impose Sommerfeld radiation boundary conditions \cite{bayliss1980radiation}. The observed data are modeled by solving \eqref{eq:waveequation} for several sources $\{q_i\}_{i=1}^{n_s}$ and sampling the resulting wavefields $u_i$ at locations $\{\vx_j\}_{j=1}^{n_r}$ and frequencies $\{\omega_k\}_{k=1}^{n_f}$, i.e.,
\[
d_{ijk} = u_i(\omega_k,\vx_j).
\]

The inverse problem is now to retrieve $m(\vx)$ from a set of observations $d_{ijk}$. This problem has been extensively studied and uniqueness results are available for a few specific cases, including layered earth-models, small perturbations around a known smooth reference medium and asymptotic versions of the problem. We refer to \cite{Symes2009} for an extensive overview. 

\subsection{Discretization and optimization}
A common approach to the inverse problem is to first 
discretize and solve the wave equation for forward modelling 
and to subsequently formulate a finite-dimensional data-fitting problem.

A fintie-difference discretization of \eqref{eq:waveequation} on an $N$-point grid with absorbing boundary conditions leads to a sparse system of equations
\[
A(\omega,\vm)\vu = \vq,
\]
where $A \in \mathbb{C}^{N\times N}$ is structurally symmetric 
% \rmrk{also with various boundary conditions?} 
and indefinite. Various dispersion-minimizing finite-difference stencils have been proposed in the literature \cite{shin1998frequency,jo1996optimal,tam1993dispersion} and several absorbing boundary conditions are described in \cite{chew1996perfectly,clayton1980absorbing}.

The system of equations can be solved by a decomposition of $A$ such as the LU (lower and upper triangular) decomposition. The advantage of this direct approach is that, once the decomposition has been performed, the system can be solved efficiently for multiple sources by forward and backward substitution \cite{marfurt1984accuracy,min2003weighted}. The direct approach with nested dissection \cite{george} is efficient for 2-D problems \cite{jo1996optimal,vstekl1998accurate,hustedt2004mixed}. However, the time and memory complexities of LU factorization and its limited scalability on large-scale distributed memory platforms prevent its application to large-scale 3-D problems that may involve more than 10 million unknowns \cite{operto20073d}. Helmholtz-specific factorization methods have been developed \cite{Wang2011}, but these are only suitable when the computational cost can be amortized over many sources. 

Iterative solvers provide an alternative approach for solving the Helmholtz equation \cite{riyanti2006new,plessix2006review,erlangga2008iterative,knibbe}. These iterative solvers are usually implemented with Krylov subspace methods \cite{saad2003iterative} that are preconditioned by solving a damped Helmholtz equation. The solution of the damped equation is computed efficiently with a multigrid method. The main advantage of the iterative approach is the low memory requirement, although the main drawback results from the difficulty to design an efficient preconditioner because $A$ is indefinite \cite{Ernst2012a}. Also, the iterative
scheme has to be started again for each source.

Organizing the observations in a vector $\vd\in\mathbb{C}^{M}$ with $M = n_s n_r n_f$, we introduce the forward operator
\[
\vd = F(\vm).
\]
Application of the forward operator on a given model $\vm$ involves the solution of $n_f n_s$ Helmholtz equations, including the LU-decomposition, and constitutes the main computational cost of FWI. 

The conventional least-squares formulation of FWI can now be expressed as
\begin{equation}
\label{eq:NLLS}
\min_{\vm} \, \left\{ f(\vm) = \textstyle{\frac{1}{2}}\| F(\vm) - \vd \|_2^2  \right\}.
\end{equation}
This optimization problem is typically solved with a Newton-like algorithm \cite{pratt1998gauss}:
\begin{align*}
\vm_{k+1} = \vm_{k} - \lambda_k \mH_k^{-1} \nabla f (\vm_{k}),
\end{align*}
where $\lambda_k$ is the step size and $H_k$ denotes (an approximation of) the Hessian of $f$ at iteration $k$. The gradient of the objective is given by 
\begin{align*}
\nabla f (\vm) = J(\vm)^* (F(\vm) - \vd),
\end{align*}
where $J(\vm)$ is the Jacobian of $F$ and $J^*$ denotes its adjoint. In practice, the Jacobian matrix is never formed explicitly but its action is computed using the \emph{adjoint-state method} \cite{Haber2000}. This entails solving $n_s n_f$ linear systems with $A^*$ and the residual as right-hand sides. The action of the (Gauss-Newton) Hessian may be computed in a similar fashion at the cost of additional Helmholtz-solves \cite{pratt1998gauss}. This can be done cheaply in 2D if the LU factors are stored in memory \cite{mulder2002time}.

\subsection{Ill-posedness and local minima}
Unfortunately, waveform inversion is hampered by the presence of local minima in $f$ \cite{santosa89B}. In practice, this requires a good initial estimate of the parameters $\vm$.
To circumvent this problem, the data can be inverted in a multi-scale fashion, starting at some lowest frequency available in the data and using the inversion result to initialize a next pass at a higher frequency \cite{Bunks1995}. The low frequencies are important in reconstructing the large-scale variations in the model, while high frequencies fill in the details. 
Other multi-scale continuation approaches have been proposed as well \cite{pratt1990inverse,Bunks1995,sirgue2004efficient,brossier2009seismic}. 
All of these will fail, however, when the initial guess does not explain the observations well enough for the lowest available frequency. 

Many alternative formulations have been proposed that depart from the usual data-fitting approach \cite{symes91,plessix99,vanLeeuwen2013Penalty1,Symes2014,Bharadwaj2016}. While these approaches can to some extent mitigate the non-linearity of the problem, they do not solve the inherent ill-posedness of the problem. This means that some features of $\vm$ are simply not recoverable, regardless of the method we use to estimate the parameters. To address this problem, we need to add regularization.

\subsection{Regularization}
We distinguish two types of regularization: \emph{implicit} regularization, where we add a penalty $\rho(\vm)$ to the objective in \eqref{eq:NLLS} to penalize unwanted features, and \emph{explicit} regularization, where we expand $\vm$ in an appropriate basis that contains only the features we desire. For example, when we expect the model to vary smoothly, we can penalize the second derivative by a penalty term
\[
\rho(\vm) = \| \mL \vm\|_2^2,
\]
with $L$ the discrete Laplace operator. Alternatively, we can choose a representation of the form
\[
\vm = \mB \va,
\]
where $B$ consists of smooth basis functions such as B-splines. This type of regularization works well when the scales of the model are separable, since we can invert for a smoothly varying velocity from low-frequency data \cite{Bunks1995}. 

In some geological settings, however, the scales do \emph{not} separate, and we need to find an alternative form of regularization.  If we expect our model to have strong discontinuities, a popular choice is a Total-Variation (TV) regularization with $\rho(\vm) = \| \mD \vm\|_1$, with $\mD$ a discrete gradient operator \cite{rudin1992nonlinear,Lin2014}. A natural basis is hard to define in this case. Alternatively, we can regularize the model by imposing the constraint $\| \mD \vm\|_1 \leq \tau$ \cite{Esser2015}. 
 
A disadvantage is that TV regularization acts globally and causes the model to be blocky everywhere. Nevertheless, some promising results have been obtained recently \cite{Esser2015,asnaashari2013regularized,anagaw2012edge}.

\section{Level-Set Method}
\label{levelset}

\begin{figure}[!ht]
  \centering
    \includegraphics[width=0.5\columnwidth]{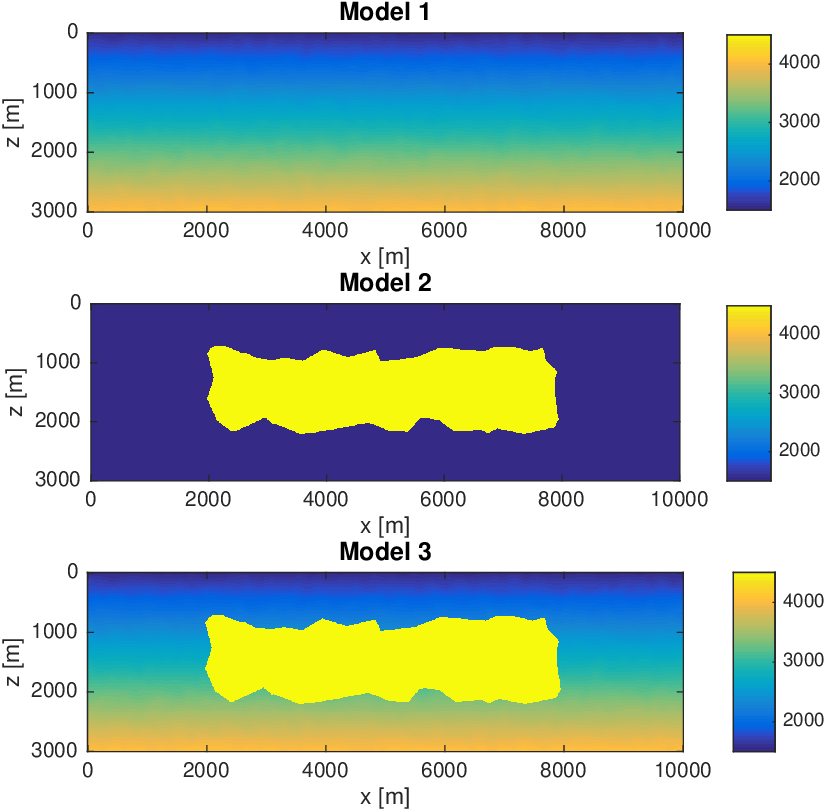}
    \caption{Model 1: smooth velocity variation (sediment). Model 2: circular blob with higher velocity (salt). Model 3: combination of smooth variation and blocky model.}
    \label{fig:models}
\end{figure}

Here, we investigate an alternative to TV regularization and propose a mixed representation for the particular case of salt bodies. We represent $m(\vx)$ as being constant in a certain region and continuously varying elsewhere. We start again from the continuous formulation of the inverse problem and derive a finite-dimensional optimization problem analogous to \eqref{eq:NLLS}.
We represent $m$ as
\[
m(\vx) = 
\left\{
\begin{array}{cc}
m_1    & \mbox{if}\ \vx \in \Omega,\\
m_0(\vx) & \mbox{otherwise}.\\ 
\end{array}
\right.
\]
Here, $\Omega$ indicates the salt-body, $m_1$ is the constant value of the model parameter inside the salt body and $m_0(\vx)$ denotes the spatially varying parameters in the sediment. Fig.~\ref{fig:models} sketches three different models, representing the smooth variation, blocky structure and a combination of both. Model 1 is a typical sediment structure, while model 2 represents the salt geometry. We generally expect a seismic velocity distribution similar to model 3, combining model 1 \& 2.

%__________________________________________________
%__________________________________________________

We can represent the model formally as 
\[
m(\vx) = [1 - \chi_{\Omega}(\vx)]m_0(\vx) + \chi_{\Omega}(\vx)m_1,
\]
where $\chi_{\Omega}(\vx)$ is the indicator function of $\Omega$. The inverse problem now consists of finding the set $\Omega$ and the model parameters $m_0(\vx)$ and $m_1$. 

The basic idea behind the level-set method is to represent the domain $\Omega$ through a level-set function as $\Omega = \{\vx\, |\, \phi(\vx) \geq 0\}$ \cite{osher2001level,mulder1992}. This then leads us to represent the indicator function as $\chi_{\Omega}(\vx) = h(\phi(\vx))$, where $h$ is the Heaviside function $h(s) = (1 + \sign (s))/2$. 

To be able to compute sensitivities, one typically uses a smooth approximation of the Heaviside function. A common choice is
\[
h_{\epsilon}(s) = \frac{1}{1 + e^{-s/\epsilon}},
\]
where $h_{\epsilon} \rightarrow h$ as $\epsilon \rightarrow 0$. This function has the nice property that its derivative is everywhere non-zero. A disadvantage is that, in order to accurately represent the indicator function, the level-set function $\phi$ will need to tend to $\pm \infty$. This induces very steep gradients in $\phi$ around the boundary of the level-set. In turn, these steep gradients in $\phi$ require that we pick a proportionally large $\epsilon$ to remain sensitive to changes in the level-set. This suggest that we pick $\epsilon$ in accordance with the (maximum) gradient of $\phi$. We will get back to this observation in Section \ref{shape}. 

To avoid some of these issues, we use a smooth Heaviside defined by
\[
h_{\epsilon}(s) = \begin{cases}
    0       & \quad \text{if } s < -\epsilon, \\
    \frac{1}{2}\bigr[1 + \frac{s}{\epsilon} + \frac{1}{\pi}\sin(\frac{\pi s}{\epsilon}) \bigr]  & \quad \text{if } -\epsilon \leq s \leq \epsilon, \\
    1  & \quad \text{if } s > \epsilon.\\
  \end{cases}
\]
To avoid getting trapped in the region where $h_{\epsilon}' = 0$, we again have to pick $\epsilon$ in accordance with the (maximum) gradient of $\phi$. A practical heuristic to pick $\epsilon$ and adapt it to the current $\phi$ will be discussed in Section \ref{shape}.

The level-set method was originally introduced for tracking regions in fluid flow applications, providing a natural way to evolve the level-set by solving a Hamilton-Jacobi equation \cite{osher1988fronts}. 
In applications like FWI, it is not obvious how to update $\phi$ away from the boundary of $\Omega$, because $h_{\epsilon}'(x)$ quickly tends to zero. The problem of finding the level-set function is ill-posed.

%====================================================

\subsection{A parametric level-set approach}
\label{pls}
To mitigate this, we adopt a method proposed in \cite{aghasi2011parametric} and represent the level-set function with a finite set of radial basis functions (RBFs):
\[
\phi(\vx) = \sum_{i=1}^{L} \alpha_i\Psi(\beta\|\vx - \vxi_i\|_2),
\]
where $\Psi(r)$ is a RBF, $\vxi_i$ are the nodes and $\beta$ is a scaling parameter. The choice of RBF will be discussed in more detail in the next subsection.

Discretization on an $N$-point grid leads to the so-called RBF-kernel matrix $\mK \in \mathbb{R}^{N\times L}$ with elements $k_{ij} = \Psi(\beta\|\vx_i - \vxi_j\|_2)$, allowing us to represent the parameters as
%__________________________________________________
\begin{equation}
\label{eq:model}
\vm(\vm_0, m_1,\valpha) = \vm_0 \odot \left(\vone - h_{\epsilon}\left(\mK\valpha\right)\right) + m_1 h_{\epsilon}\left(\mK\valpha\right),
\end{equation}
%__________________________________________________
where $\odot$ represents element-wise multiplication, also known as Hadamard product.
We can now define the corresponding optimization problem for the parametric level-set approach, for fixed $\vm_0, m_1$, as
%__________________________________________________
\begin{equation}
\label{eq:PLS}
\min_{\valpha} \{ \widetilde{f}(\valpha) = \textstyle{\frac{1}{2}}\| F(\vm(\valpha)) - \vd \|_2^2\}.
\end{equation}
%__________________________________________________
The gradient of this objective is given by
%__________________________________________________
\begin{equation}
\label{eq:PLSgrad}
\nabla \widetilde{f}(\valpha) = \left(\frac{\partial \vm}{\partial\valpha}\right)^T\nabla f(\vm(\valpha)) ,
\end{equation}
where 
\[ 
\frac{\partial \vm}{\partial\valpha} =  \mathrm{diag} \left\{ (m_1 \vone - \vm_0) \odot h_{\epsilon}' (\mK\valpha) \right\} \mK .
\]
%__________________________________________________

%====================================================

\subsection{Radial basis functions}
Radial basis functions are a means to approximate smooth multivariate functions. They have been extensively studied in the context of
the interpolation of scattered data in high dimensions
and for meshless methods \cite{zhang2000meshless,buhmann2003radial,powell1987radial}.
RBFs are classified into two main types, global RBFs, which have infinite support, and compactly supported RBFs. Next, we discuss some relevant properties when we consider approximating a given smooth function $\phi$ using RBFs.

\subsubsection{Global RBFs}
\begin{figure}[!t]
\centering
\centering
\subfloat[]{\includegraphics[width=3.0in]{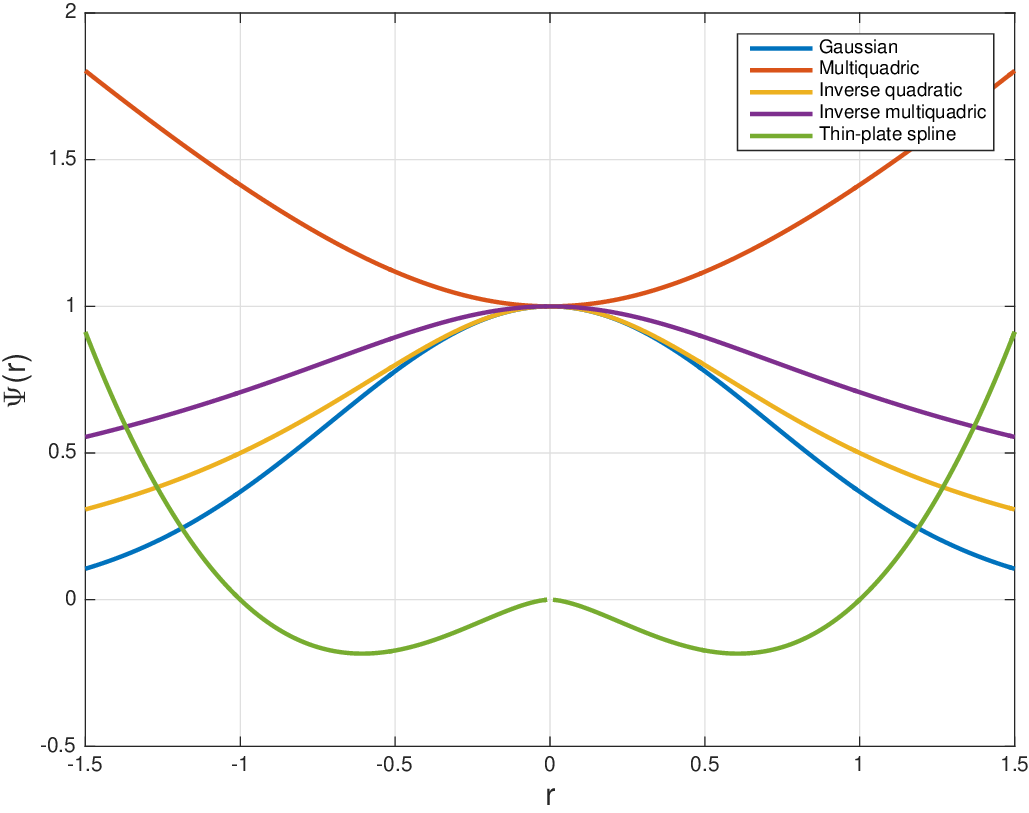}\label{fig:global}}
\hfil
\subfloat[]{\includegraphics[width=3.0in]{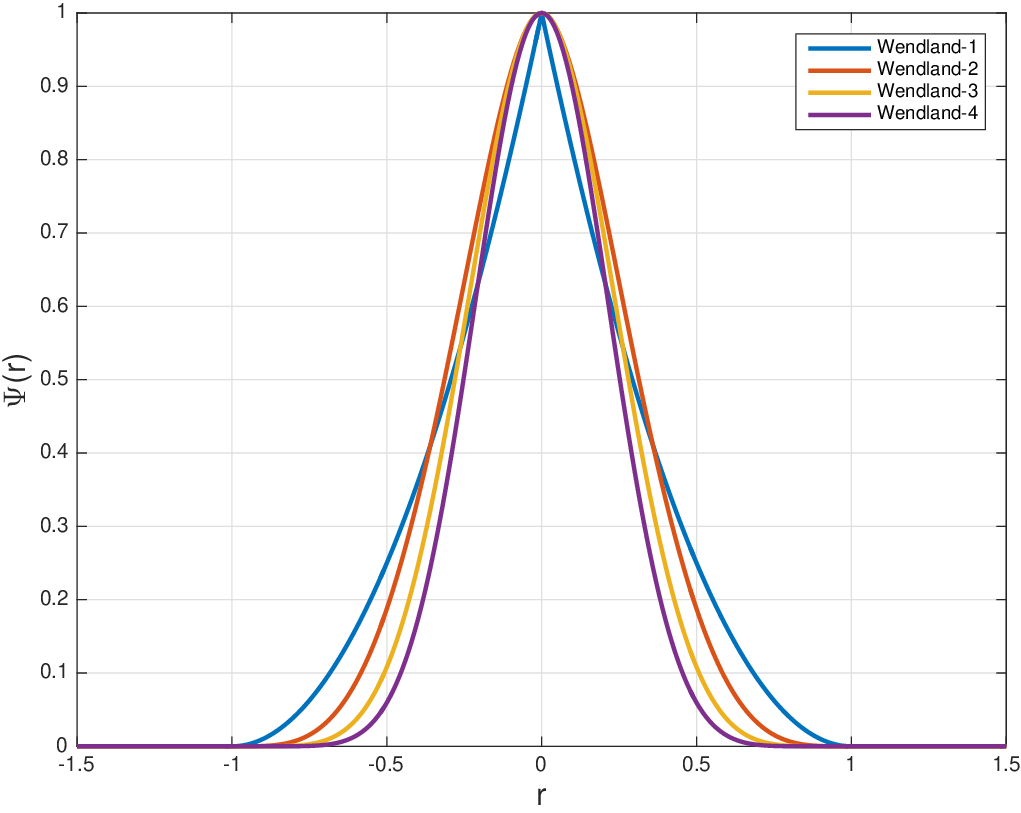}\label{fig:wendland}}
\caption{(a) Various types of global RBF. (b) Various types of Wendland RBFs.}
\end{figure}

Global RBFs have infinite support and hence the RBF kernel matrix $K$ is dense.
An overview of several common global RBFs is given 
in Table \ref{table:globalrbf} and their radial behavior is shown in Fig.~\ref{fig:global}.
For multiquadric and thin-plate spline,
the kernel matrix is positive definite \cite{micchelli1984interpolation}. 
Among the advantages of global RBFs are
(1) highly accurate and often exponentially convergent,
(2) easy applicable to high-dimensional problems,
(3) meshless in the approximation of multivariate scattered data and 
(4) numerical accuracy is easily improved by adding more points in regions with large gradients.

However, the corresponding interpolation matrix is dense and ill-conditioned and therefore sensitive to the shape parameter. As a result, the application
of traditional RBF interpolation to large-scale problems is computationally expensive.

%__________________________________________________
\begin{table}[!t]
\caption{Global RBFs.}
\centering
\begin{tabular}{l | r}
\toprule
Name & $\Psi(r)$ \\
\midrule
Gaussian & $\exp(-r^2)$ \\
Multiquadric & $\sqrt{1+r^2}$ \\
Inverse multiquadric & $\frac{1}{\sqrt{1+r^2}}$ \\
Inverse quadratic & $\frac{1}{1+r^2}$ \\
Thin-plate spline  &  $r^2 \ln(r)$ \\
\bottomrule
\end{tabular}
\label{table:globalrbf}
\end{table}
%__________________________________________________
%__________________________________________________

\subsubsection{Compactly supported RBF}\label{sec:wendland}
these result in a sparse, positive definite, and generally better conditioned kernel matrix \cite{wendland1995piecewise}. However, the order of approximation is usually less than with global RBFs. Table~\ref{table:compactrbf} provides
an overview of some common compactly supported RBFs. Fig.~\ref{fig:wendland} presents their radial behavior.

%__________________________________________________
\begin{table}
\caption{Wendland compactly-supported RBFs. Here, $k$ denotes the order of smoothness (i.e., $\phi$ is k times continuously differentiable).}
\centering
\begin{tabular}{c | lc}
\toprule
Name & $\Psi(r)$&$k$\\
\midrule
Wendland-1 & $(1-r)^2_+$ & 0 \rule{0pt}{2.6ex}\\
Wendland-2 & $(1-r)^4_+ (4r + 1)$ & 2 \rule{0pt}{2.6ex}\\
Wendland-3 & $(1-r)^6_+(35r^2 + 18r + 3)$ & 4 \rule{0pt}{2.6ex}\\
Wendland-4 & $(1-r)^8_+(32r^3 +25r^2 +8r+1)$ & 6 \rule{0pt}{2.6ex}\\
\bottomrule
\end{tabular}
\label{table:compactrbf}.
\end{table}
%__________________________________________________

\subsection{Shape representation}
\label{shape}
To determine which type of RBF is most suitable for our purpose, we study how well we can represent typical salt bodies with various RBFs. These salt models are discretized on a cartesian grid with grid spacing $h$. We choose the nodes of the RBFs on a cartesian grid with a larger grid spacing $h_r = 5 h$ and normalize the scale parameter for the compactly supported RBFs with $\beta = \frac{1}{\gamma h_r}$. 

\begin{figure*}[!t]
\centering
\subfloat[]{\includegraphics[width=1.5in]{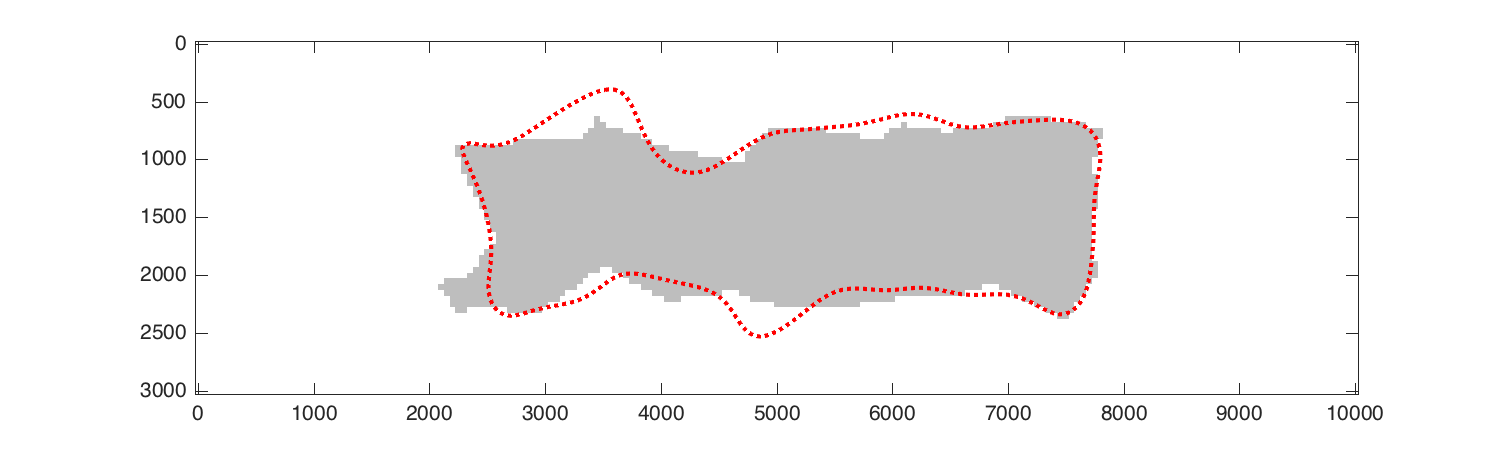}}
\hfil
\subfloat[]{\includegraphics[width=1.5in]{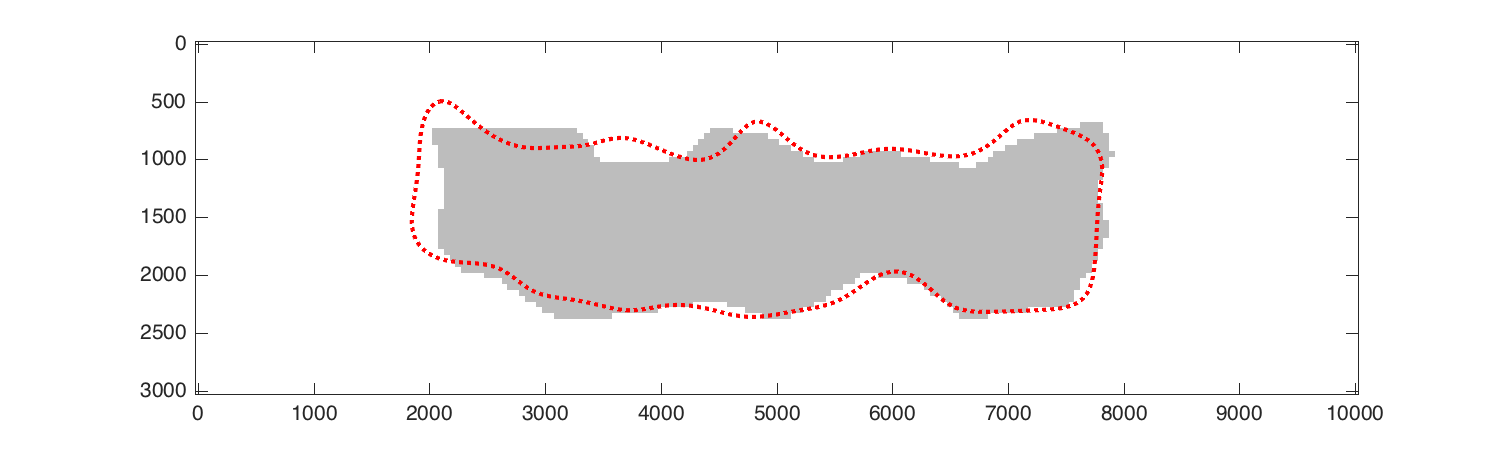}}
\hfil 
\subfloat[]{\includegraphics[width=1.5in]{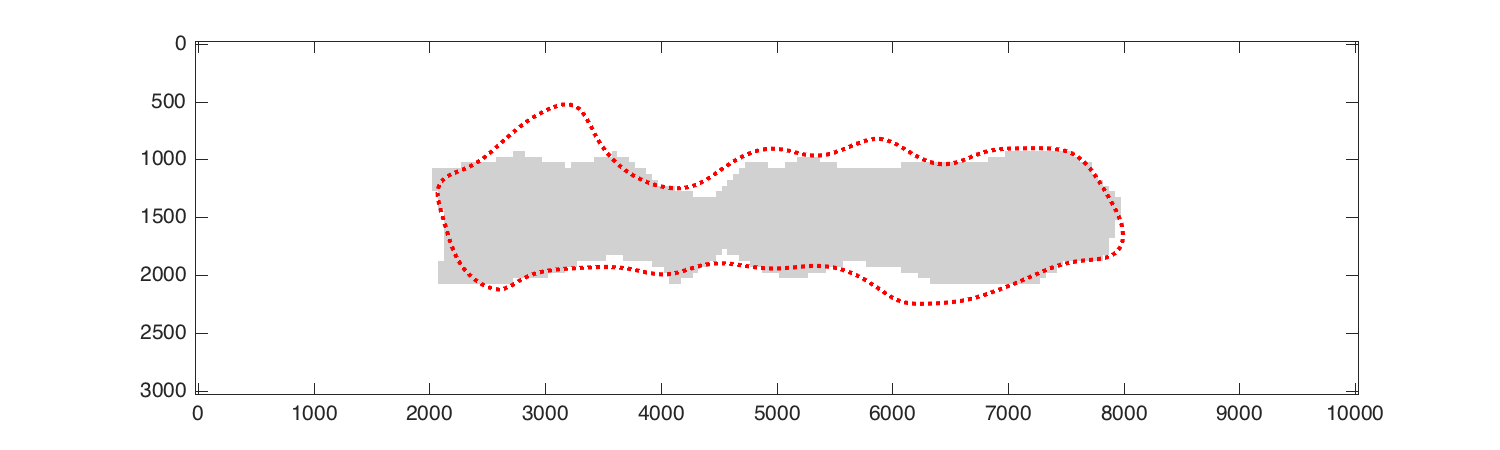}}
\hfil 
\subfloat[]{\includegraphics[width=1.5in]{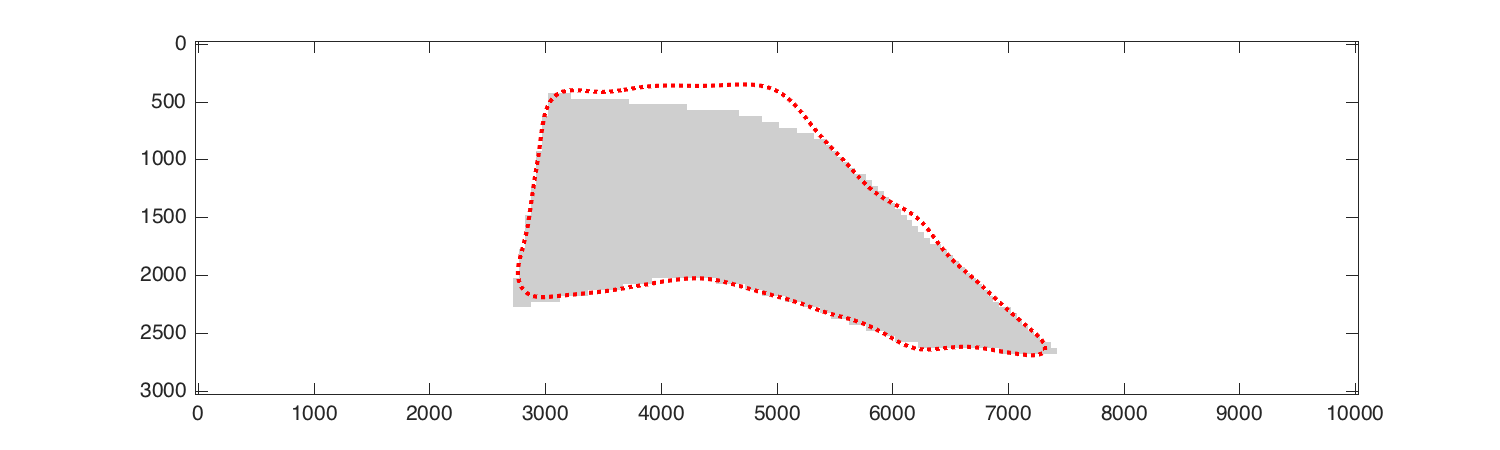}}
\hfil
\subfloat[]{\includegraphics[width=1.5in]{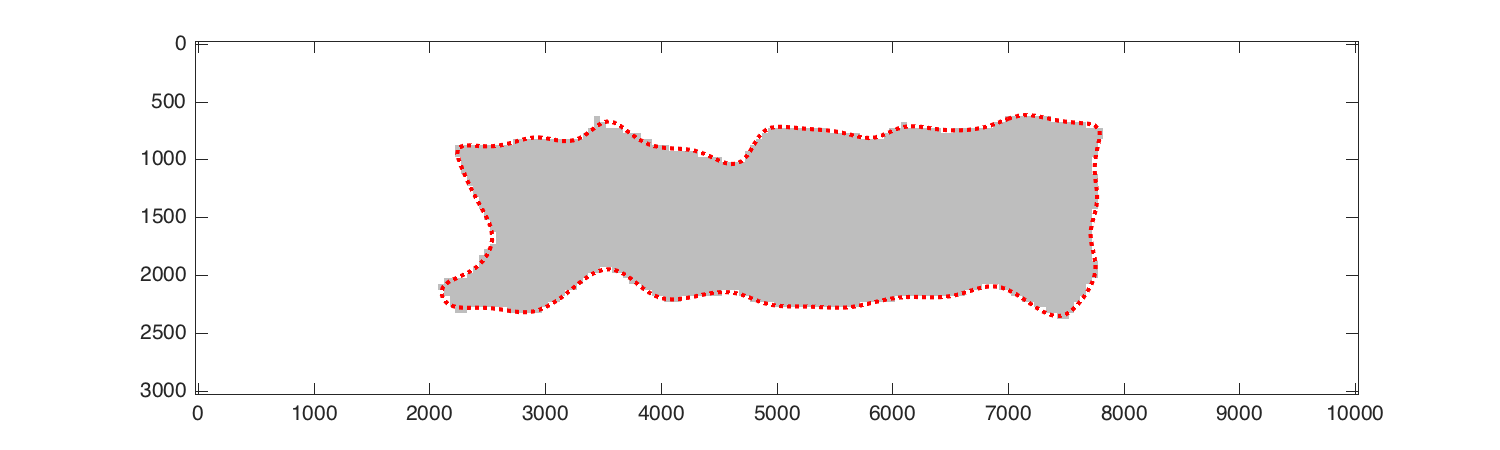}}
\hfil
\subfloat[]{\includegraphics[width=1.5in]{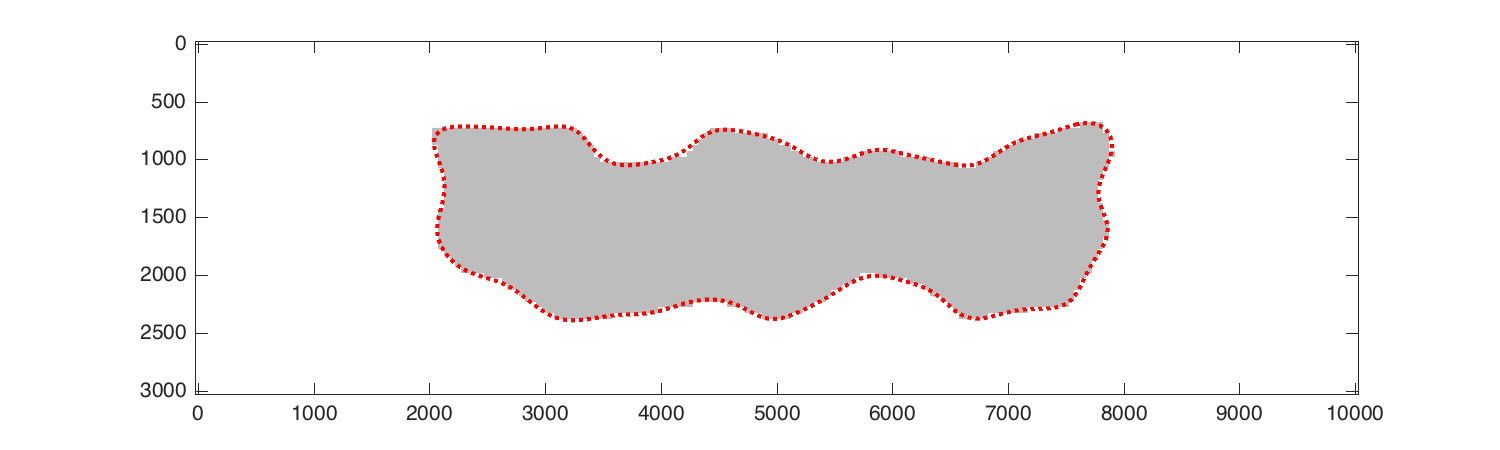}}
\hfil 
\subfloat[]{\includegraphics[width=1.5in]{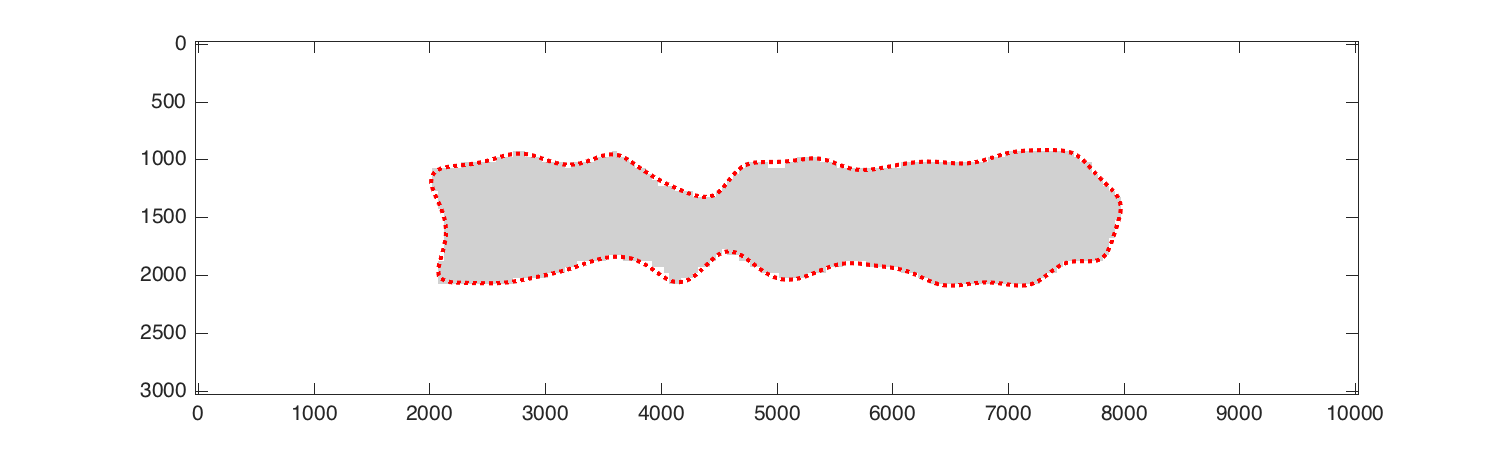}}
\hfil 
\subfloat[]{\includegraphics[width=1.5in]{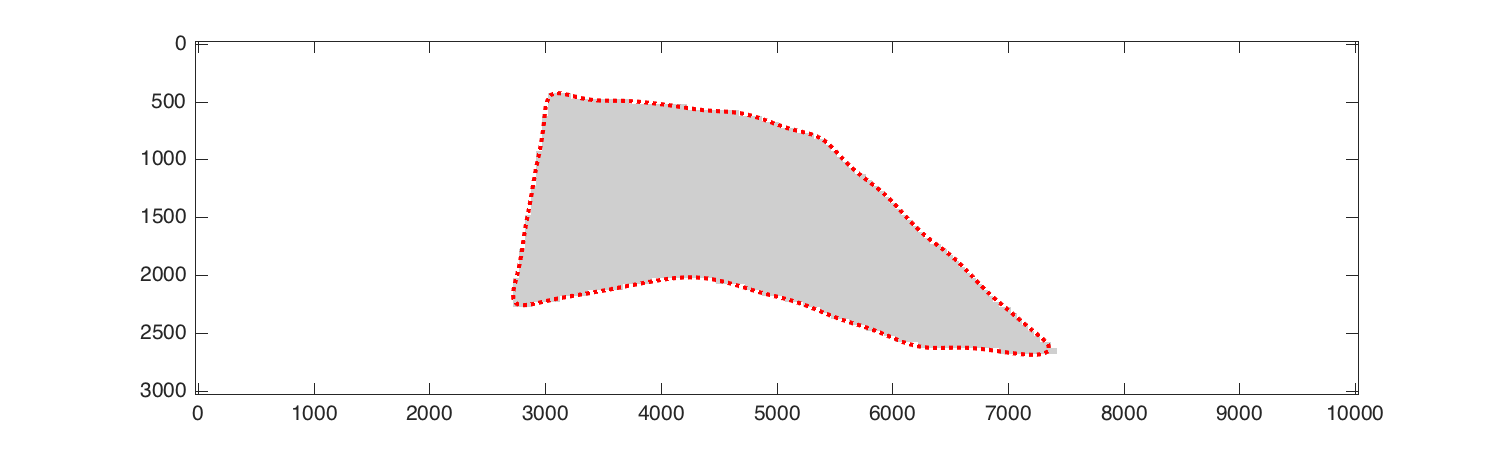}}
\caption{Improved reconstructions from new formulation of $\epsilon$ (red dotted line denotes reconstructed level-set). }
\label{fig:heaviOvsN}
\end{figure*}

We determine the coefficients $\valpha$ by solving 
\[
\min_{\valpha} \textstyle{\frac{1}{2}}\|h_{\epsilon}(\mK_{\gamma}\valpha) - \vm\|_2^2,
\]
with a L-BFGS method. Results for the Wendland-4 RBF with $\gamma = 4$ and $\epsilon = 10^{-1}$ are shown in Fig.~\ref{fig:heaviOvsN}(a--d). The lower-order Wendland RBFs gave a less good approximation. Results with the global RBFs are similar to those of the compact ones. 

As noted earlier in Section~\ref{levelset}, we need to pick $\epsilon$ in accordance with the (current) level-set function in order to have optimal sensitivity. We propose to choose $\epsilon$ adaptively based on the current level-set function as
%__________________________________________________
\begin{equation}
\label{eq:heavieps}
\epsilon = \tfrac{1}{2}\kappa \left[\max(K_\gamma\valpha) - \min(\mK_\gamma\valpha) \right].
\end{equation}
%__________________________________________________
This choice of $\epsilon$ with $\kappa = 10^{-1}$ 
produced the results in Fig.~\ref{fig:heaviOvsN}(e--h). 

%__________________________________________________
%__________________________________________________

%====================================================

\subsection{Algorithm}
Algorithm~\ref{alg:basic} summarizes
parametric level-set full waveform inversion (PLS-FWI).

%__________________________________________________
\begin{algorithm}[H]
 \caption{PLS-FWI Basic Algorithm}
 \label{alg:basic}
 \begin{algorithmic}[1]
 \REQUIRE Data $\vd$, forward modeling operator $\mathbf{F}$, estimate of background parameter $\vm_0$, salt parameter $m_1$, RBF kernel matrix $\mK_\gamma$, initial estimate of weights $\valpha$
 \ENSURE model $\vm$
 \STATE $\kappa \leftarrow 0.1$
 \FOR {$j = 1$ to itermax }
 \STATE compute Heaviside $\epsilon$ from equation~\eqref{eq:heavieps}
 \STATE compute misfit $\widetilde{f}(\valpha)$ from equation~\eqref{eq:PLS} and gradient $\nabla \widetilde{f} (\valpha)$ from equation~\eqref{eq:PLSgrad}
 \STATE $\valpha \leftarrow\valpha + \lambda \widetilde{\mH}^{-1} \nabla \widetilde{f}(\valpha)$
 \ENDFOR 
 \STATE $\epsilon \leftarrow 0$
 \STATE compute $\vm$ from equation~\ref{eq:model}.
 \end{algorithmic}
\end{algorithm}
%__________________________________________________

Algorithm~\ref{alg:multiscale} outlines the multi-scale approach of PLS-FWI. We reduce $\kappa$ for Heaviside $\epsilon$ after every frequency band to decrease the size of level-set boundary. The idea is to start optimization with large boundary (high $\kappa$) and small initial level-set to capture large sensitivity, allowing for large updates. Decreasing the level-set boundary then provides sharper images.

%__________________________________________________
\begin{algorithm}[H]
 \caption{PLS-FWI Multi-scale Algorithm}
 \label{alg:multiscale}
 \begin{algorithmic}[1]
 \REQUIRE Data $\vd \in \mathbb{R}^{n_{s} \times n_{r} \times n_f}$, forward modeling operator $F$, estimate of background parameter $\vm_0$, salt parameter $m_1$, RBF kernel matrix $\mK_\gamma$, initial estimate of weights $\valpha$
 \ENSURE model $\vm$
 \STATE $\kappa \leftarrow 0.1$ 
 \FOR {$i = 1$ to $n_f$}
 \STATE Compute $\valpha$ from PLS-FWI basic algorithm~\ref{alg:basic}
 \STATE $\kappa \leftarrow 0.8 \kappa $
 \ENDFOR
 \STATE $\epsilon \leftarrow 0$
 \STATE compute $\vm$ from equation~\ref{eq:model}.
 \end{algorithmic}
\end{algorithm}
%__________________________________________________

\section{Results}
\label{results}
\begin{figure*}[!t]
\centering
\subfloat[]{\includegraphics[width=3.0in]{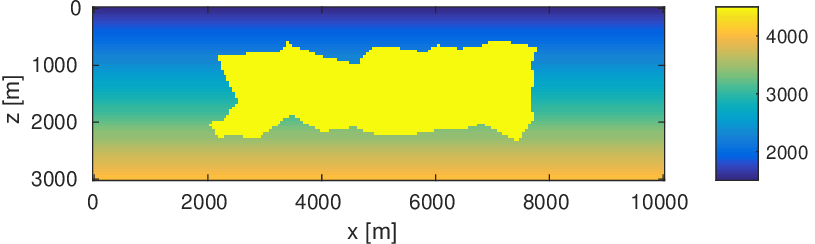}\label{fig:m1true}}
\hfil
\subfloat[]{\includegraphics[width=3.0in]{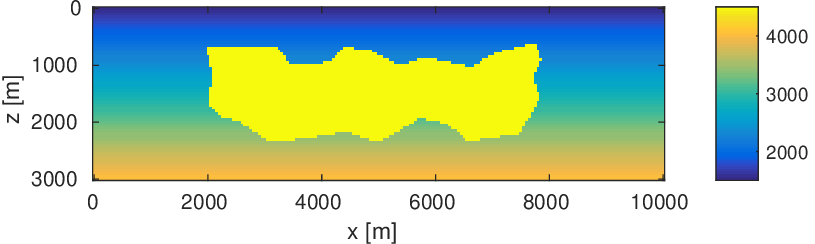}\label{fig:m2true}}
\hfil 
\subfloat[]{\includegraphics[width=3.0in]{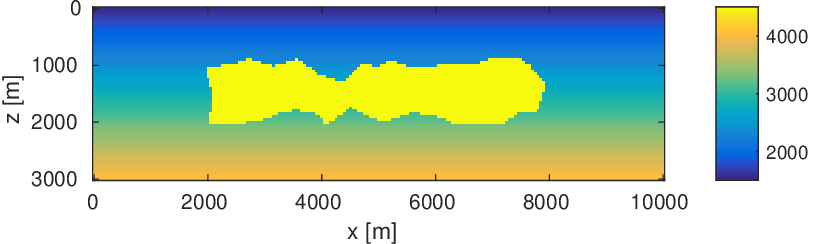}\label{fig:m3true}}
\hfil 
\subfloat[]{\includegraphics[width=3.0in]{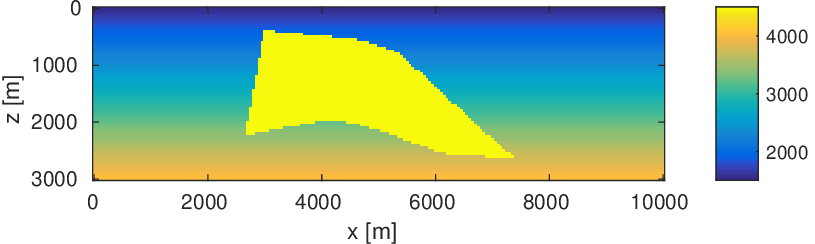}\label{fig:m4true}}
\caption{Velocity models created for the inversion test. They are referred to as model A~\protect\subref{fig:m1true}, model B~\protect\subref{fig:m2true}, model C~\protect\subref{fig:m3true} and model D~\protect\subref{fig:m4true}. }
\label{fig:truemodels}
\end{figure*}

To demonstrate the applicability of the parametric level-set full waveform inversion, %(PLS-FWI),
we perform simulations on four different velocity models. These models are shown in Fig.~\ref{fig:truemodels}. Each has a background velocity increasing linearly with depth as $1500 + b z$ with $b = 0.8333$. The salt bodies in these model have a constant velocity of $4500~$m/s. We choose a grid spacing of $50~$m in each direction, providing a total of $N=201 \times 61$ grid points. 

\begin{figure}[!t]
\centering
\centering
\subfloat[]{\includegraphics[width=3.0in]{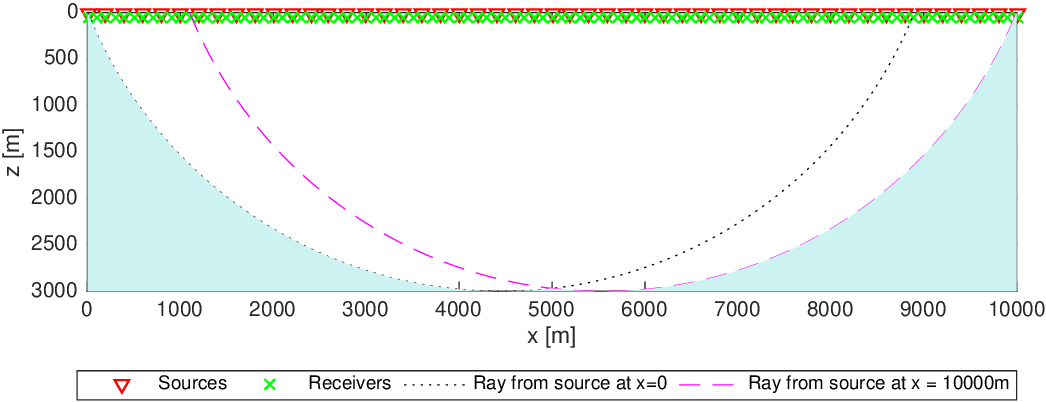}\label{fig:srcnrec}}
\hfil
\subfloat[]{\includegraphics[width=3.0in]{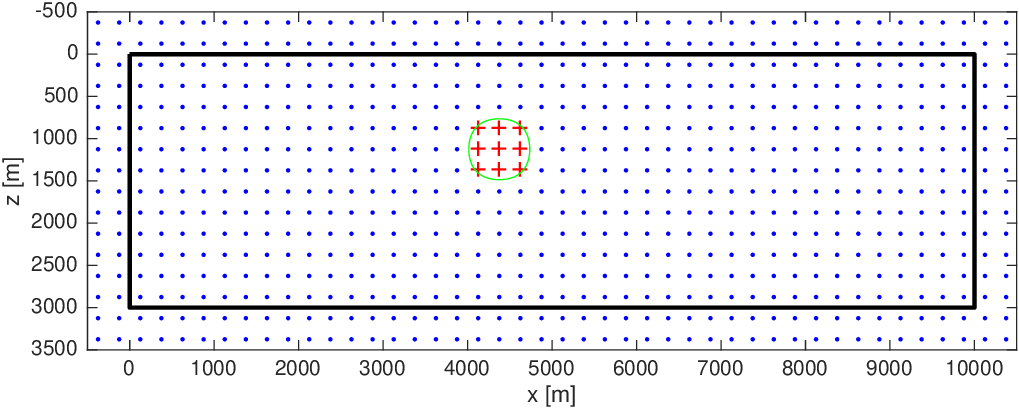}\label{fig:rbfpl}}
\caption{(a) Source and receivers for the simulation. Sources are placed on top, receivers at a depth of $50$~m. The two ray paths separate the region recoverable by inversion from the shaded non-recoverable part~\cite{kuvshinov2006exact}. (b) Placements of RBF on the computational grid (denoted by thick black line). To initialize the zero contour of the
level set (green line) that defines the salt body, a few RBFs around the center have been allocated positive values (denoted by red plusses), others are negative (denoted by blue dots). }
\end{figure}

The acquisition setup is shown in Fig.~\ref{fig:srcnrec}, with $50$ sources placed at top of the model and $100$ receivers placed at the depth of $50~$m. Also shown is the part of the model to which the data are most sensitive. Features of the model in the lower left and right corners are hard to recover because of the limited aperture.

We use a frequency-domain finite-difference code \cite{dasilva2016} to generate the data for frequencies between $2.5$ and $3.5\,$Hz with a spacing of $6.25 \times 10^{-2}\,$Hz.
The data are weighted in frequency by a Ricker wavelet with a peak frequency of $15~$Hz.
For the inversion, we select four bands, $[2.5-2.75]\,$Hz, $[2.75-3.0]\,$Hz, $[3.0-3.25]\,$Hz and $[3.25-3.5]\,$Hz, each with 4 four frequencies.

For PLS-FWI, the RBF grid has a spacing of $250\,$m in both directions. As shown in Fig.~\ref{fig:rbfpl}, two extra layers of RBF nodes are added outside the physical domain to provide flexibility in reconstructing the level-set near the boundary. We use a fourth-order Wendland RBF with $\gamma = 4$ and choose $\epsilon$ adaptively as described in Section~\ref{shape}. 

We compare FWI and PLS-FWI on noise-free and on data with white noise, with an SNR of $10\,$dB. Finally, we perform a joint reconstruction of both the salt geometry and the background model $ \vm_0 = 1500 + b z$, parametrized by the slope $b$.

For FWI, we use a projected Quasi-Newton method \cite{schmidt2009optimizing} to solve the resulting optimization problem with bound-constraints on $\vm$ to ensure that the velocity stays within the feasible range of $[1500,4500]\,$m/s.
To solve the resulting optimization problem in $\valpha$ (PLS-FWI), we rely on the L-BFGS approximation of the Hessian\cite{minfunc}. We perform 150 iterations per frequency band. For joint salt and background reconstruction with PLS-FWI, we use a bisection method to find the optimal $b$.

%__________________________________________________
%__________________________________________________
%__________________________________________________
%\begin{figure}[!t]
%\centering
%\includegraphics[width=3.0in]{rbfplacement.eps}
%\caption{Placements of RBF on the computational grid (denoted by thick black line). To initialize the zero contour of the
%level set (green line) that defines the salt body, a few RBFs around the center have been allocated positive values (denoted by red plusses), others are negative (denoted by blue dots). }
%\label{fig:rbfpl}
%\end{figure}
%__________________________________________________

\subsection{Salt geometry determination}
The initial model for conventional FWI and the background model for PLS-FWI are taken to be the same, linearly increasing velocity models as in the true ones. For PLS-FWI, we let $m_1$ correspond to the true velocity in the salt and initialize the level-set as shown in Fig.~\ref{fig:rbfpl}.

\subsubsection{Noise-free data}
\begin{figure*}[!t]
\centering
\subfloat[]{\includegraphics[width=2.5in]{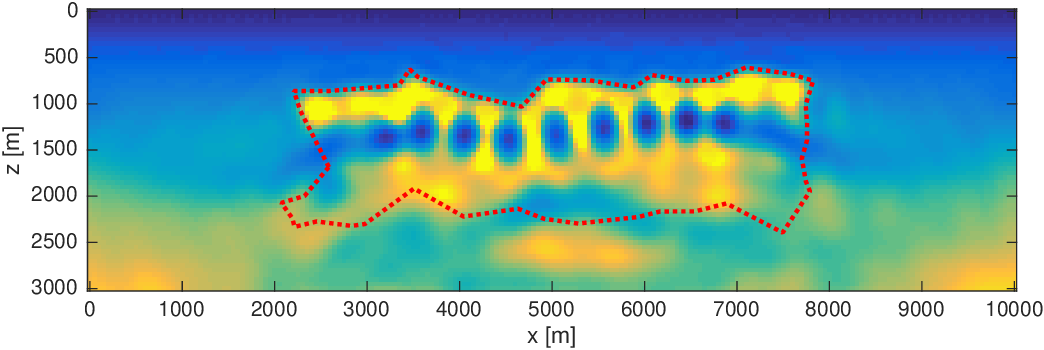}\label{fig:sPFm1c}}
\hfil 
\subfloat[]{\includegraphics[width=2.5in]{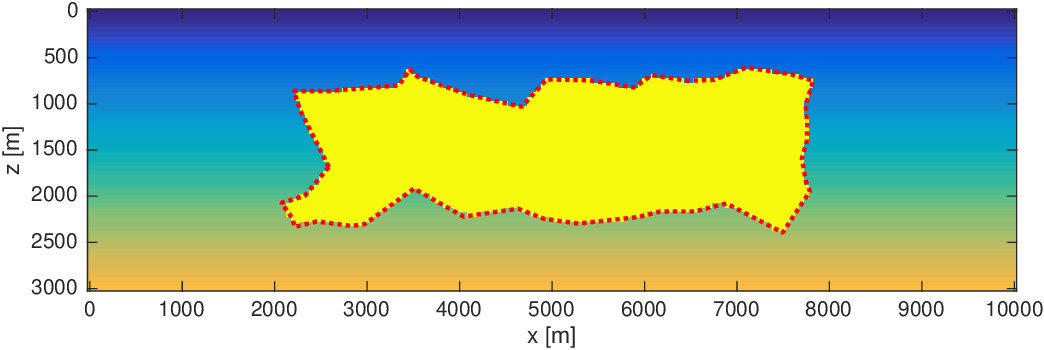}\label{fig:sPFm1p}}
\hfil
\subfloat[]{\includegraphics[width=2.5in]{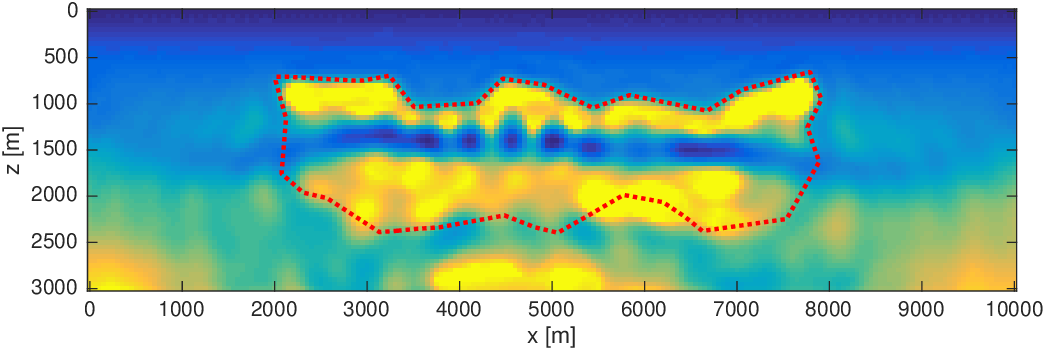}\label{fig:sPFm2c}}
\hfil
\subfloat[]{\includegraphics[width=2.5in]{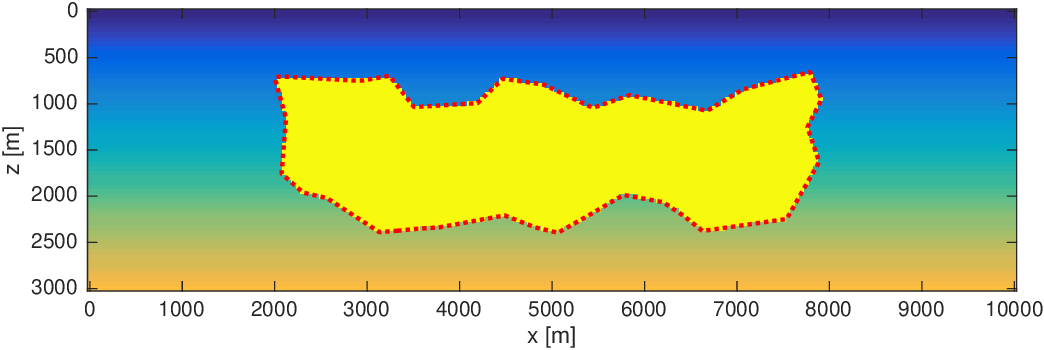}\label{fig:sPFm2p}}
\hfil
\subfloat[]{\includegraphics[width=2.5in]{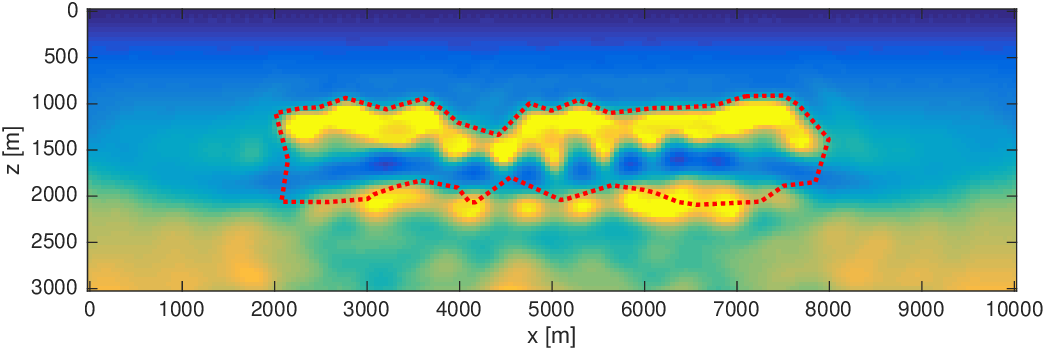}\label{fig:sPFm3c}}
\hfil
\subfloat[]{\includegraphics[width=2.5in]{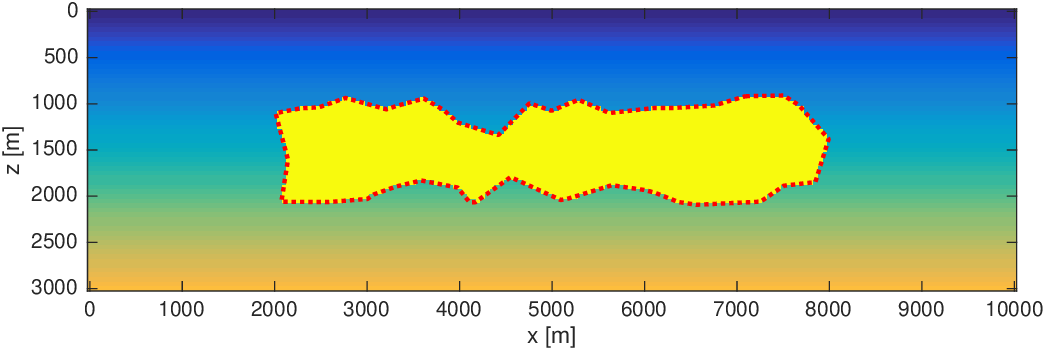}\label{fig:sPFm3p}}
\hfil
\subfloat[]{\includegraphics[width=2.5in]{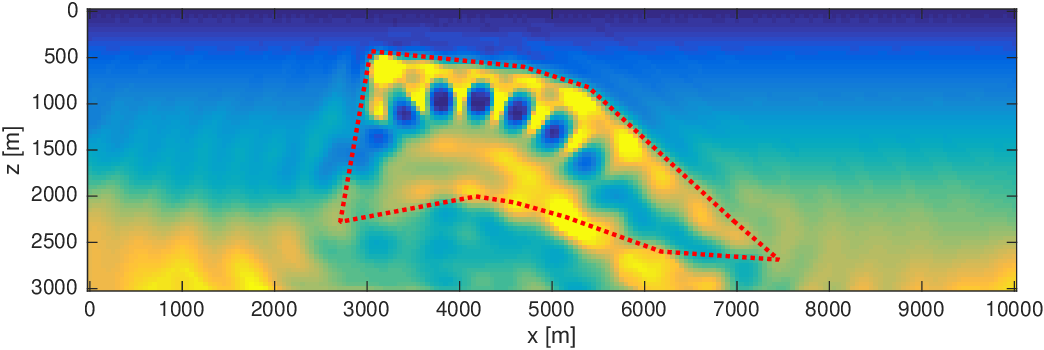}\label{fig:sPFm4c}}
\hfil
\subfloat[]{\includegraphics[width=2.5in]{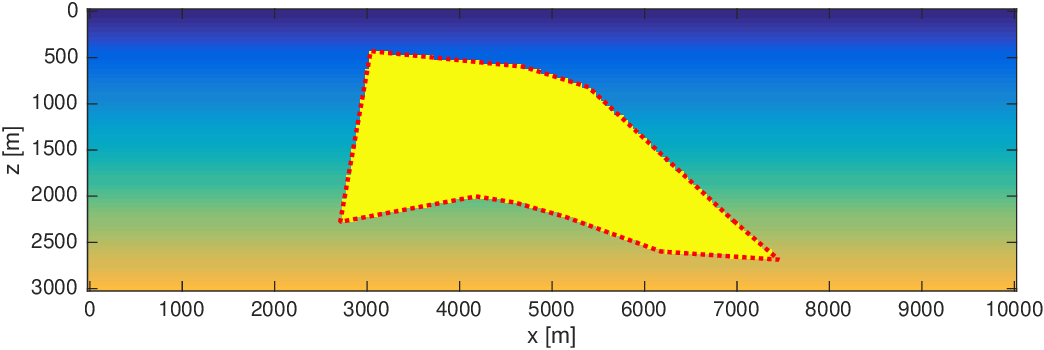}\label{fig:sPFm4p}}
\caption{Salt reconstruction with classical FWI and parametric level-set full waveform inversion (PLS-FWI) from noise free data. \protect\subref{fig:sPFm1c}, \protect\subref{fig:sPFm2c}, \protect\subref{fig:sPFm3c} and \protect\subref{fig:sPFm4c} represent models A, B, C, and D, reconstructed with classical full waveform inversion. \protect\subref{fig:sPFm1p}, \protect\subref{fig:sPFm2p}, \protect\subref{fig:sPFm3p} and \protect\subref{fig:sPFm4p} represent reconstructed model A, B, C, D using PLS-FWI. The red dotted line shows the true geometry of salt.}  
\label{fig:sPLSFWI}
\end{figure*}

Fig.~\ref{fig:sPFm1c}, \ref{fig:sPFm2c}, \ref{fig:sPFm3c} and \ref{fig:sPFm4c} show the reconstructed models using FWI. To a large extent, the results predict the top of the salt but fail to identify its proper shape. They also include some artefacts, as shown in the left bottom part of Fig.~\ref{fig:sPFm4c}. On the other hand, PLS-FWI almost completely recovers the salt geometry in each of the models, as can be seen in Fig.~\ref{fig:sPFm1p}, \ref{fig:sPFm2p}, \ref{fig:sPFm3p} and \ref{fig:sPFm4p}.

To compare the proposed method with conventional FWI, we define the error reduction factor (ERF),
\[ \text{ERF} = \frac{\| F(\vm_{recon}) - \vd \|_2}{\| F(\vm_{0}) - \vd \|_2} ,\]
where $E_{FWI}$ and $E_{PLS-FWI}$ denote the misfit errors achieved by FWI and PLS-FWI, respectively. An ERF close to the best achievable ERF ($0$ in case of noise-free data) indicates a better performance of the reconstruction method in reducing the data misfit. Table~\ref{table:misfit} shows the improvement in the data misfit with PLS-FWI over classic FWI. The data misfit is reduced by a factor $10^{-4}$ on average with the use of PLS-FWI.

To compare the reconstructions for different methods, we define a measure called the Relative Reconstruction Error
as
\[ RRE = \frac{\| \vm_{recon} - \vm_{true} \|_2}{\| \vm_{0} - \vm_{true} \|_2},\]
where $m_{recon}$ is the model reconstructed by FWI or PLS-FWI. From Table~\ref{table:misfit}, we observe that the RRE is reduced drastically with PLS-FWI.

%__________________________________________________
%__________________________________________________
\begin{table}[!t]
\renewcommand{\arraystretch}{1.3} 
\caption{Comparison of misfit and RRE values for classical FWI vs PLS-FWI}
\label{table:misfit}
\centering
\begin{tabular}{c | c c | c c }
\toprule
\multirow{2}{*}{Model} & \multicolumn{2}{c|}{ERF} & \multicolumn{2}{c}{RRE} \\
 & FWI & PLS-FWI & FWI & PLS-FWI \\
\midrule
A & $0.0281 $ & $5.8197\times 10^{-4} $ &  $0.8213$ & $0.0707$ \\  		% 2.064174e+08  &  4.273832e+06 &  6.236166e+05 (Power) & 7.343718e+09 
B & $0.0241 $ & $7.2637\times 10^{-5} $  & $0.8543$ & $0.0548$ \\   		% 1.709077e+08  &  5.142056e+05 & 6.235854e+05 (power) & 7.079092e+09
C & $0.0261 $ & $4.7127\times 10^{-4} $ & $0.9102$ & $0.0732$ \\		% 1.444293e+08  &  2.605259e+06	&  6.227972e+05 (power) & 5.528216e+09
D & $0.0339 $ & $9.0299\times 10^{-6}$ & $0.8088$ & $0.0434$ \\ 		% 2.391849e+08  &  6.368814e+04 &  6.212497e+05 (power) & 7.053037e+09 
\bottomrule
\end{tabular}
\end{table}
%__________________________________________________

\subsubsection{Noisy data}

\begin{figure}[!t]
\centering
    \includegraphics[width=0.5\columnwidth]{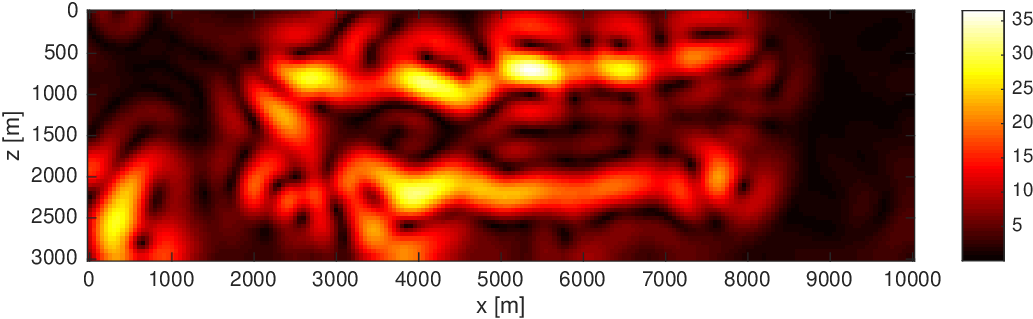}
    \caption{Gradient of the final level set function for model A with noisy data.}
     \label{fig:dphim1}
\end{figure}

Fig.~\ref{fig:sPFm1c}, \ref{fig:sPFm2c}, \ref{fig:sPFm3c} and \ref{fig:sPFm4c} show the reconstructed models using FWI while Fig.~\ref{fig:sPFm1p}, \ref{fig:sPFm2p}, 
\ref{fig:sPFm3p}, \ref{fig:sPFm4p} show the models reconstructed by PLS-FWI. 
The results are essentially the same as for the noise-free case, except for some artifacts outside the region of interest in the PLS-FWI results. Fig.~\ref{fig:dphim1} exhibit the variation of achieved level-set function for Model A. Large gradient still exists at left bottom corner which implies the difficulty to get rid of false salt geometry. 

Table~\ref{table:misfitn} shows the improvement in the data misfit with PLS-FWI over classic FWI. For noisy data, we also look at the achievable ERF which is defined as:
\[ \text{ERF}_{\text{achievable}} = \frac{\| F(\vm_{true}) - \vd \|_2}{\| F(\vm_{0}) - \vd \|_2}.\]
This quantity denotes the smallest ERF achievable by any reconstruction method. The ERF for PLS-FWI is closer to the best achievable ERF than that of FWI. With the classic approach, the relative reconstruction error is slightly affected when noise is added to data. On the other hand, RRE changes by a large factor in the presence of noise with the proposed method. The false salt geometries mainly attribute to these large changes.

%__________________________________________________
\begin{figure*}[!t]
\centering
\subfloat[]{\includegraphics[width=2.5in]{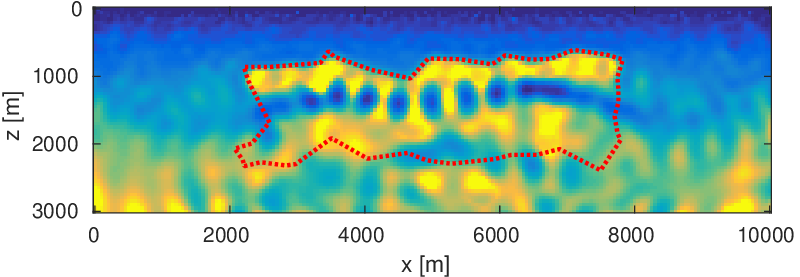}\label{fig:sPFm1cn}}
\hfil
\subfloat[]{\includegraphics[width=2.5in]{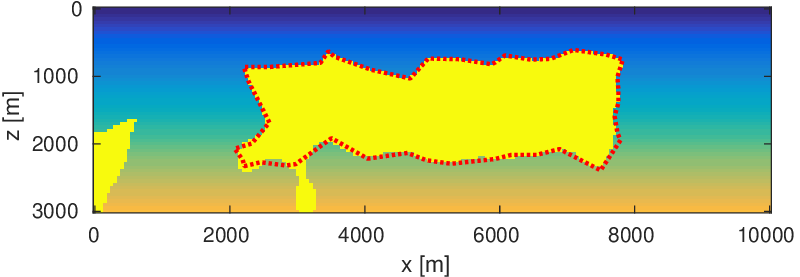}\label{fig:sPFm1pn}}
\hfil
\subfloat[]{\includegraphics[width=2.5in]{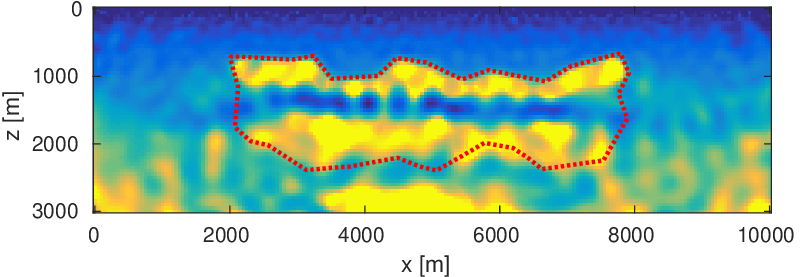}\label{fig:sPFm2cn}}
\hfil
\subfloat[]{\includegraphics[width=2.5in]{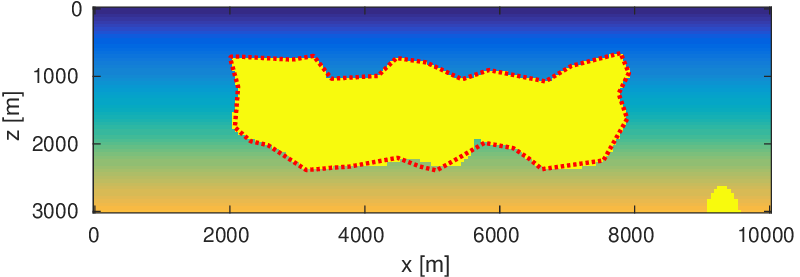}\label{fig:sPFm2pn}}
\hfil
\subfloat[]{\includegraphics[width=2.5in]{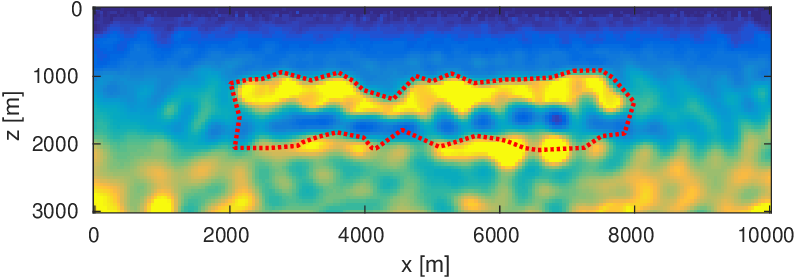}\label{fig:sPFm3cn}}
\hfil
\subfloat[]{\includegraphics[width=2.5in]{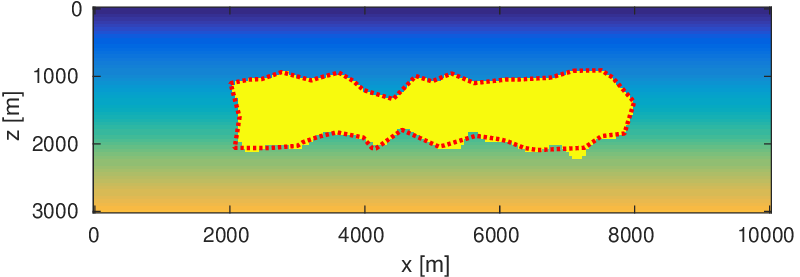}\label{fig:sPFm3pn}}
\hfil
\subfloat[]{\includegraphics[width=2.5in]{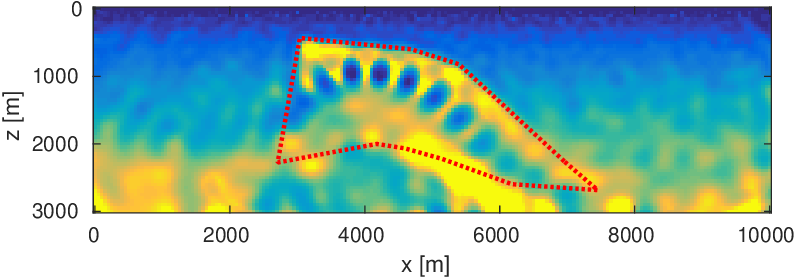}\label{fig:sPFm4cn}}
\hfil 
\subfloat[]{\includegraphics[width=2.5in]{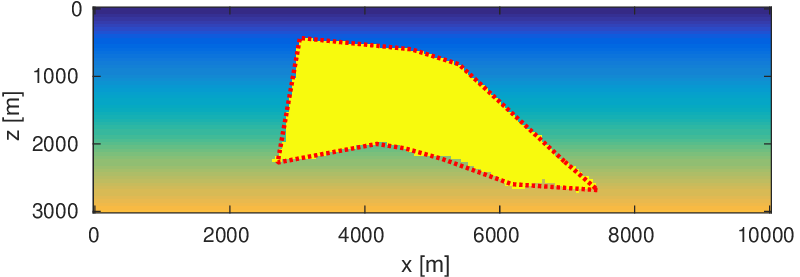}\label{fig:sPFm4pn}}
\caption{Salt reconstruction with classic FWI and Parametric level-set full waveform inversion (PLS-FWI) with noisy data of SNR $10~dB$. \protect\subref{fig:sPFm1c}, \protect\subref{fig:sPFm2c}, \protect\subref{fig:sPFm3c}, \protect\subref{fig:sPFm4c} represent reconstructed model A, B, C, D with classic full waveform inversion respectively. \protect\subref{fig:sPFm1p}, \protect\subref{fig:sPFm2p}, \protect\subref{fig:sPFm3p}, \protect\subref{fig:sPFm4p} represent reconstructed model A, B, C, D with PLS-FWI respectively. Red dotted line shows the true geometry of salt.}
\label{fig:sPLSFWIn}
\end{figure*}
%__________________________________________________
%__________________________________________________
\begin{table}[!t]
\renewcommand{\arraystretch}{1.3}
 \caption{Comparison of misfit and RRE values for classic FWI vs. PLS-FWI (noisy data).}
\label{table:misfitn}
\centering
\begin{tabular}{c | c c c | c c}
\toprule
\multirow{2}{*}{Model} & \multicolumn{3}{c|}{ERF} & \multicolumn{2}{c}{RRE} \\
& FWI & PLS-FWI & achievable & FWI & PLS-FWI \\
\midrule
A & $ 0.5425 $ & $ 0.5254 $ & $0.5246$ & $0.8432$ & $0.2437$  \\ % 8.337037e+09 & 8.105888e+09  ||  8.401848e+09 & 8.137201e+09 &  6.731762e+05 (power) & 1.548678e+10 & 8.123641e+09
B & $ 0.5499 $ & $ 0.5347 $ & $0.5340$ & $0.8548$ & $0.1508$ \\ % 8.278382e+09 & 8.045434e+09  ||  8.374004e+09 & 8.141737e+09 &  6.731682e+05 (power) & 1.522775e+10 & 8.132366e+09 
C & $ 0.6090$ & $ 0.5958$ & $ 0.5951 $ & $0.9199$ & $0.1817$\\ % 8.357810e+09 & 8.078406e+09  ||  8.317237e+09 & 8.136656e+09 &  6.723124e+05 (power) &  1.365679e+10 &  8.127694e+09 
D & $ 0.5530$ & $ 0.5325 $ & $ 0.5321$ & $0.8239$ & $0.1382$ \\ % 8.229564e+09 & 7.962226e+09  ||  8.320599e+09 & 8.012409e+09 &  6.701038e+05 (power) &  1.504648e+10  &  8.005674e+09 
\bottomrule
\end{tabular}
\end{table}
%__________________________________________________

\subsection{Simultaneous reconstruction of salt and background}
Next, we jointly reconstruct the salt geometry and the background, parametrized by $b$. For each frequency band, we first estimate the optimal $b$, for fixed $\valpha$ and setting $\epsilon = 0$, using a bisection method and subsequently estimate the salt geometry with PLS-FWI (Algorithm~\ref{alg:basic}).

\begin{figure*}[!t]
\centering
\subfloat[]{\includegraphics[width=2.5in]{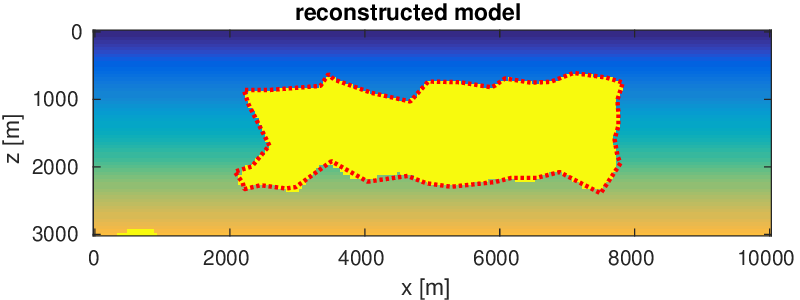}\label{fig:bsPFm1n}}
\hfil
\subfloat[]{\includegraphics[width=2.5in]{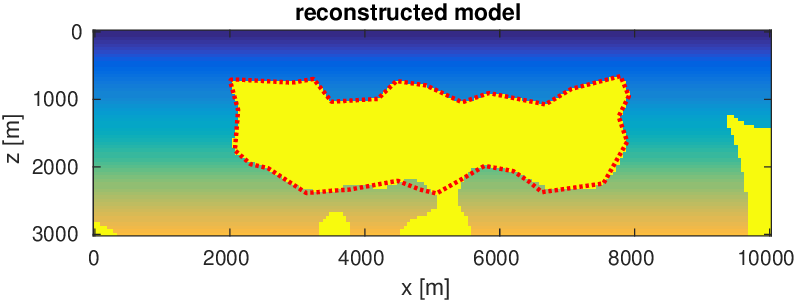}\label{fig:bsPFm2n}}
\hfil
\subfloat[]{\includegraphics[width=2.5in]{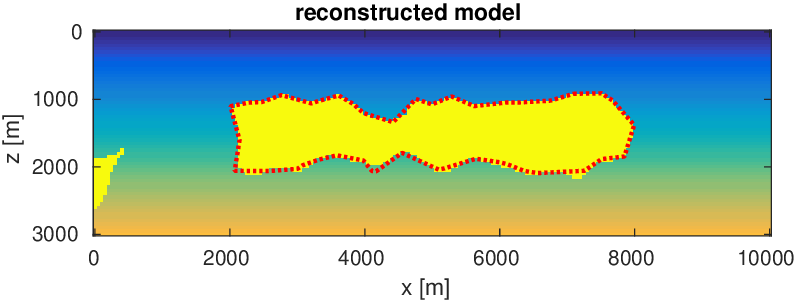}\label{fig:bsPFm3n}}
\hfil
\subfloat[]{\includegraphics[width=2.5in]{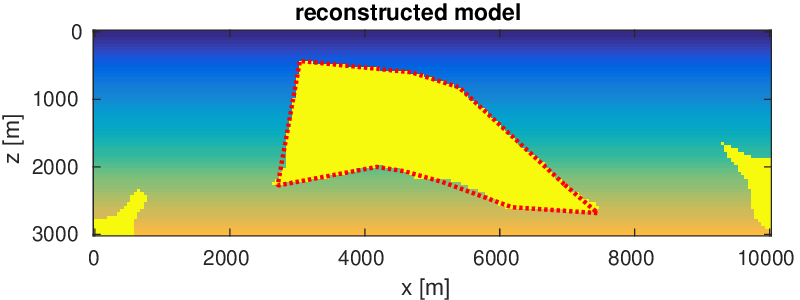}\label{fig:bsPFm4n}}
\label{fig:bsPLSFWIn}
\caption{Simultaneous reconstruction of salt and background with Parametric level-set full waveform inversion (PLS-FWI) on noisy data with SNR $10~dB$. \protect\subref{fig:bsPFm1n}, \protect\subref{fig:bsPFm2n}, \protect\subref{fig:bsPFm3n}, \protect\subref{fig:bsPFm4n} represent reconstructed model A, B, C, D respectively. Red dotted line shows the true geometry of salt.}
\end{figure*}

Fig.~\ref{fig:bsPFm1n}, \ref{fig:bsPFm2n}, \ref{fig:bsPFm3n} and \ref{fig:bsPFm4n} show the reconstructed models for noisy data. The salt geometry is accurately predicted, but contains a few artifacts. 
Table~\ref{table:bvaluen} indicates the predicted values of $b$ in each of the models, which are very close to true value of $b=0.8333$. 

%__________________________________________________
%__________________________________________________
\begin{table}[!t]
\renewcommand{\arraystretch}{1.3}
\caption{Recovered values of $b$ for different models (noisy data).}
\label{table:bvaluen}
\centering
\begin{tabular}{l | r r r r}
\toprule
Model & A & B & C & D \\
\midrule
$b$ & $0.8351$ & $0.8352$ & $0.8345$ & $0.8333$ \\
\bottomrule  
\end{tabular}
\end{table}
%__________________________________________________

\section{Discussions and Conclusion}
\label{conclusions}

Accurately determining the geometry of subsurface salt bodies from seismic data is a difficult problem. When casting the inverse problem into a non-linear data-fitting problem, both the presence of local minima and the ill-posedness of the problem prevent accurate recovery of the salt-geometry. We have investigated the application of a parametric level-set method to address this problem. We represent the Earth model as a continuously varying background with an embedded salt body.The salt geometry is described by the zero contour of a level-set function, which in turn is represented with a relatively small number of radial basis functions. This formulation includes some additional parameters such as the width of the basis functions and the smoothness of the Heaviside function. The latter is of particular importance as it controls the sensitivity to changes in the salt geometry.
We propose a robust algorithm that adaptively chooses the required smoothness parameter and tested the method on a suite of idealized Earth models with different salt geometries. For a fixed and accurate background model, the level-set method is shown to give superior estimates of the salt geometry and is stable against a moderate amount of noise. Additional results demonstrate that is feasible to jointly estimate the background and the salt geometry.

To further develop the method as a viable alternative to conventional full-waveform inversion, tests on more realistic Earth models are needed. In particular, the joint estimation of the background model and salt geometry needs to be investigated further. Even when representing the level-set function with a finite basis, there are many level-set functions that result in the same salt geometry. To address this issue, additional regularization is needed. An often-used approach is to re-initialize the level-set function by solving an Eikonal equation. In the parametric framework, this can be included by adding the discretized Eikonal equation as a regularization term. 

In this paper, we fixed RBF grid \emph{a priori}. For very complex salt-geometries, this may no longer be feasible as it would require too many nodes for their accurate representation. An adaptive choice of the RBF grid may address this problem but it is not obvious how to refine the grid.

\section*{Acknowledgment}

This work is part of the Industrial Partnership Programme (IPP) `Computational sciences for energy research' of the Foundation for Fundamental Research on Matter (FOM), which is part of the Netherlands Organisation for Scientific Research (NWO). This research programme is co-financed by Shell Global Solutions International B.V. The second author is financially supported by the Netherlands Organisation for Scientific Research (NWO) as part of research programme 613.009.032. We acknowledge Seismic Laboratory for Modeling and Imaging (SLIM) at University of British Columbia (UBC) for providing computational software. They thank Eldad Haber for stimulating discussions and pointing out the applicability of the parametric level-set method.

\end{document}